\documentclass[prd,12pt,aps,superscriptaddress,double-spaced,floatfix,showkeys,showpacs]{revtex4}
\usepackage{graphicx,amsmath,bbm,bm,array,subfigure}
\usepackage[linktocpage=true,colorlinks=true,linkcolor=blue,citecolor=blue]{hyperref}

\begin{document}

\title {Electrical and thermal conductivities of hot and dense hadronic matter}
\author{Guruprasad Kadam }
\email{guruprasadkadam18@gmail.com}
\affiliation{Department of Physics,
	Shivaji University, Kolhapur,
	Maharashtra 416004, India}
\author{Hiranmaya Mishra}
\email{hm@prl.res.in}
\affiliation{Theory Division, Physical Research Laboratory,
 Navarangpura, Ahmedabad - 380 009, India}
 \author{Lata Thakur\footnote{present address : National Institute of Science Education and Research, HBNI, 752050 Odisha, India}}
\email{latathakur@niser.ac.in}
\affiliation{Theory Division, Physical Research Laboratory,
 Navarangpura, Ahmedabad - 380 009, India}


\def\be{\begin{equation}}
\def\ee{\end{equation}}
\def\bearr{\begin{eqnarray}}
\def\eearr{\end{eqnarray}}
\def\zbf#1{{\bf {#1}}}
\def\bfm#1{\mbox{\boldmath $#1$}}
\def\hf{\frac{1}{2}}
\def\sl{\hspace{-0.15cm}/}
\def\omit#1{_{\!\rlap{$\scriptscriptstyle \backslash$}
{\scriptscriptstyle #1}}}
\def\vec#1{\mathchoice
        {\mbox{\boldmath $#1$}}
        {\mbox{\boldmath $#1$}}
        {\mbox{\boldmath $\scriptstyle #1$}}
        {\mbox{\boldmath $\scriptscriptstyle #1$}}
}

\begin{abstract}
We estimate the electrical and thermal conductivities of hot and dense hadronic matter in the relaxation time approximation of the Boltzmann equation. We estimate the thermodynamical quantities of hot and dense hadronic matter within the ambit of the excluded volume hadron resonance gas model.  The relaxation time for all the hadrons is estimated assuming the constant cross section with uniform as well as mass dependent hard-core radius. We compare our results with various  existing results. Finally we give an estimate of electrical and thermal conductivities in the context of heavy ion collision experiments. 
\end{abstract}

\maketitle

\section{Introduction}

Transport coefficients of hot and dense matter are one of the challenging contemporary research interests particularly
in the field of strong interaction physics. These are interesting quantities for several reasons. For many 
physical systems, through their dependences on system parameters like temperature, chemical potential
can reveal the location of the phase transition in the phase diagram. In the context of heavy ion collisions (HICs), the matter produced in
the fireball after a collision, with quarks and gluons degrees of freedom behaves like a strongly interacting liquid 
with a small shear viscosity it expands, cools and undergoes a crossover transition to hadronic degrees of freedom which 
finally free stream to the detector. One of the successful descriptions of such an evolution is through dissipative relativistic hydrodynamics ~\cite{Gale:2013da,Schenke:2011qd,Shen:2012vn,Kolb:2003dz,Teaney:2000cw,DelZanna:2013eua,Karpenko:2013wva,
Holopainen:2011hq,Jaiswal:2015mxa} and transport simulations~\cite{Xu:2004mz,Bouras:2010hm,Bouras:2012mh,Fochler:2010wn,Wesp:2011yy,Uphoff:2012gb,Greif:2013bb,Danielewicz:1984ww}. 
 Finite but small shear viscosity ($\eta$) to entropy ($s$) ratio is necessary to explain the flow data
~\cite{Gyulassy:2004zy,Csernai:2006zz}. The smallness of this ratio 
$\frac{\eta}{s}$ and its connection to the conjectured Kovtun-Son-Starinets bound 
of $\frac{\eta}{s}=
\frac{1}{4\pi}$ obtained using
AdS/CFT correspondence \cite{Kovtun:2004de} has motivated many theoretical investigations of this ratio to understand and derive rigorously
from a microscopic theory  
 ~\cite{Gavin:1985ph,Prakash:1993bt,Dobado:2003wr,Itakura:2007mx,Chen:2007xe,Dobado:2009ek,Demir:2008tr, Puglisi:2014pda,Thakur:2017hfc}. 
The other viscosity coefficient $\zeta$ has also been realized to be important to be included the dissipative hydrodynamics. During the expansion of the fireball, when the temperature approaches the critical temperature $\zeta$ can be large and give rise to different interesting
phenomena like cavitation when the pressure vanishes and hydrodynamic description breaks down \cite{Rajagopal:2009yw, Bhatt:2011kr}. The
effect of bulk viscosity on the particle spectra and flow coefficients have been investigated \cite{Monnai:2009ad,Denicol:2009am,Dusling:2011fd}
while the interplay of shear and bulk viscosity coefficients have been studied in Refs. \cite{Song:2009rh,Noronha-Hostler:2013gga,Noronha-Hostler:2014dqa}. The coefficient of bulk viscosity has been estimated for both the hadronic and the partonic systems
~\cite{Dobado:2001jf, Davesne:1995ms,Kharzeev:2007wb,FernandezFraile:2009mi,Chen:2007kx,NoronhaHostler:2008ju,Sasaki:2008um,FernandezFraile:2008vu,Dobado:2012zf,Ozvenchuk:2012kh,Gangopadhyaya:2016jrj,Chakraborty:2010fr,Sasaki:2008fg}.
In the case of non central and asymmetric HICs, 
a large magnetic field as well as electric field is expected to be produced \cite{Tuchin:2013ie,Voronyuk:2011jd}. The event by event analysis for the Relativistic Heavy Ion Collision system indicates the generation of the
magnetic field on the order of $eB\simeq m_\pi^2$ as well as the electric field $eE\simeq m_\pi^2$. 
The strong magnetic field so produced has exciting possibilities of observing CP violating effects known as the chiral magnetic and chiral
vortical effects. Apart from these, there have been other dynamical manifestations of such strong fields on other observables, like an increase in
the elliptical flow coefficient. 
However, all these interesting and important effects in off central heavy ion collisions require that a resoanbly strong
magnetic field survives for at least several fermi proper time. 
Initially it was thought that the magnetic field decays rapidly 
after the collision\cite{khar}. It was later pointed out that the rapid decrease in the magnetic field 
leads to 
induced electric current that slows down the decrease of the magnetic field and satisfies a diffusion equation \cite{tuchin,arpanajit}. The  crucial
parameter that goes in to the estimation of the time 
scale of this diffusion is the electrical conductivity of the medium 
$\sigma_{el}$. The time evolution of the magnetic field in relativistic heavy ion collisions is still an open question.
This requires  a proper estimate of the electrical conductivity of the medium as well as solutions of magnetohydrodynamic
equations which need further investigation \cite{tuchin,amreddy}.
This apart, $\sigma_{el}$ also enters in the hydrodynamic evolution, where charge relaxation also plays an 
important role. This coefficient
influences significantly the soft photon production~\cite{Yin:2013kya} as well as low mass 
dilepton enhancement \cite{Akamatsu:2011mw}.  

 Several groups have studied the electrical conductivity, including the chiral perturbation theory~\cite{Fukushima:2008xe}, the numerical solution of the Boltzmann equation~\cite{Greif:2014oia, Puglisi:2014sha}, holography~\cite{Finazzo:2013efa}, transport models ~\cite{Cassing:2013iz,Steinert:2013fza}, Dyson Schwinger calculations~\cite{Qin:2013aaa}, a dynamical quasiparticle model~\cite{Marty:2013ita,Berrehrah:2014ysa}, a quasiparticle model~\cite{Srivastava:2015via,Thakur:2017hfc}, the effective fugacity quasiparticle model~\cite{Mitra:2016zdw}, and lattice gauge theory~\cite{Gupta:2004,Aarts:2007wj,Buividovich:2010tn,Ding:2010ga,Burnier:2012ts,Brandt:2012jc,Amato:2013naa}. All these studies aim at the value of $\sigma_{\text{el}}$ in the QGP phase, but some of these do extend below the transition
 temperature towards the hadron gas. 
Despite the importance of  electrical conductivity, it has rarely been studied in the  literature for the hadronic phase.
 Recently, $\sigma_{\text{el}}$ has been investigated  for a pion gas~\cite{Fernandez-Fraile2006} and for hot hadron gas~\cite{Greif:2016skc,Ghosh:2016yvt,Samanta:2017ohm}. It has also been studied in the framework of the
 Polyakov-Nambu-Jona-Lasinio model~\cite{Saha:2017xjq} and the Polyakov-Quark-Meson (PQM) model~\cite{Singha:2017jmq}.

 The transport coefficient that plays an important role in the hydrodynamic evolution at finite baryon densities is the coefficient of thermal
conductivity ($ \kappa $). The effects of thermal conductivity in the relativistic hydrodynamics has been recently emphasized in Refs.~\cite{Denicol:2012vq, Kapusta:2012zb}. The thermal conduction which involves relative flow of energy and baryon number, vanishes at zero baryon density. However,
for situations, where, e.g. the pion number is conserved, particularly at low temperatures, heat conductivity can be sustained by pions
which themselves have zero baryon number ~\cite{Gavin:1985ph}. 
Recently, thermal conductivity has been studied for pionic medium by different groups~\cite{Gavin:1985ph,Mitra:2014dia,Ghosh:2015mba,Dobado:2007cv,Prakash:1993bt,Davesne:1995ms}.
 The heat conductivity was also
 obtained using the Kubo formula~\cite{FernandezFraile:2009mi,Fraile2,SS_AHEP} and Nambu-Jona-Lasinio (NJL) Model~\cite{Deb:2016myz,Marty:2013ita}. Heat conductivity has been investigated recently in a transport model~\cite{Greif:2013bb} and a PQM coupling model~\cite{Abhishek:2017pkp}. 
 
  
We might note here that it is of practical as well as fundamental importance to estimate the transport coefficients also in the
hadronic phase to distinguish the signatures of QGP matter and hadronic matter.
These coefficients can be estimated directly within QCD using Kubo formulation. However, as QCD is strongly coupled 
for the energies accessible in heavy ion collision experiments, the task is very nontrivial. First principle calculations 
like lattice QCD simulation are also challenging and are limited to equilibrium properties at small chemical potentials.
These coefficients therefore have been estimated within various effective models for strong interaction as well as various approximations in the estimation. 

In the present work we intend to estimate the coefficients of electrical conductivity and thermal conductivity for the
hadronic phase within the ambit of a hadron resonance gas model (HRGM). The HRG model, which successfully  describes the hadronic phase
with the  multiplicities of particle abundances of various hadrons in heavy ion collisions 
\cite{BraunMunzinger:1995bp,Yen:1998pa,Becattini:2000jw}, is assumed to be a free gas of all observed hadrons and their resonances
treated as point particles. As shown in Ref.\cite{Dashen:1969ep}, this is a reasonable way to include attractive interaction among hadrons.  
Apart from hadronic multiplicities, this model has been used to estimate viscosity coefficients \cite{Gorenstein:2007mw, NoronhaHostler:2012ug,Tiwari:2011km,NoronhaHostler:2008ju,Kadam:2015xsa} as well as the study of
fluctuations in conserved charges in HIC experiments \cite{Bhattacharyya:2015zka,Garg:2013ata}. However, the simple HRGM misses the repulsive 
interactions among hadrons, the existence of  which is already known from nucleon-nucleon scattering experiments.  Such repulsive interactions 
can be implemented  via an excluded volume approximation whereby the volume available for the hadrons to move is reduced by the volume they occupy~\cite{Hagedorn:1980kb,Hagedorn:1982qh,Gorenstein:1981fa}. This HRGM  
with excluded volume (EHRGM)~\cite{Gorenstein:2007mw,Rischke:1991ke,Kapusta:1982qd,Albright:2014gva,Kadam:2015xsa} corrections has been found to be in good agreement with lattice QCD results up to temperature, $T\sim 140$MeV.
The model has also been used to estimate the viscosity coefficients using relaxation time approximation for solving the relativistic Boltzmann kinetic equation~\cite{Kadam:2015xsa}. We use here a similar approximation to
estimate the electrical and thermal conductivities of hadronic matter.\\
 
We organize the paper as follows. In Sec.~\ref{secII}, we recapitulate the excluded volume hadron 
resonance gas model.  In Sec.~\ref{secIII}, we compute the electrical and thermal
conductivities using the relativistic Boltzmann equation in relaxation time approximation relevant for multicomponent hadronic medium. In 
Sec.~\ref{secIV}, we calculate the relaxation time in the limit of isotropic constant scattering cross section for the hadrons.
In Sec.~\ref{secV}, we discuss our results and finally in Sec.~\ref{secVI}, we summarize findings of the present investigation.

\section{Excluded volume hadron resonance gas model}
\label{secII}
As we have discussed in the Introduction, hadrons cannot be considered as point particles. The repulsive interactions can be taken into account between hadrons via an excluded volume approximation or van der Waals treatment. 
The thermodynamic pressure is related to the partition function as
\be
P_{\text{id}}=T \lim_{V \rightarrow \infty}\frac{\text{ln} Z_{\text{id}}(T,\mu,V)}{V},
\ee
where $T$ is temperature, $\mu$ is chemical potential, and $V$ is volume of the system.
In thermodynamically consistent excluded volume formulation, one can obtain the transcendental equation for the pressure as~\cite{Rischke:1991ke, Yen:1997rv}

\be
P^{EV}(T,\mu)=P^{\text{id}}(T,\tilde{\mu}),
\label{prexcl}
\ee
where $\tilde{\mu}=\mu-vP^{EV}(T,\mu)$ is an effective chemical potential with $ v $ as the parameter 
corresponding to proper volume of the particle.  At high temperature and low densities this prescription is equivalent to multiplying a
suppression factor of $\text{exp}(-vP^{EV}/T)$ to the pressure in the Boltzmann approximation. Therefore, the pressure  in excluded volume hadron resonance gas model becomes

\be
P^{EV}(T,\mu)=e^{\frac{-vP^{EV}(T,\mu)}{T}}P^{{id}}(T,\mu),
\ee
where $P^{id}$ in Boltzmann approximation can be written as

\be
P^{{id}}(T,\mu)=\sum_{a}\frac{g_{a}}{2\pi^{2}}m_{a}^{2}T^{2}K_{2}\bigg(\frac{m_{a}}{T}\bigg)\text{cosh}\bigg(\frac{\mu}{T}\bigg),
\ee
where $g_{a}$ is the degeneracy of $a$th hadron species. 
Other thermodynamical quantities can be readily obtained from Eq. (\ref{prexcl}) by taking appropriate derivatives.  The number density, energy density, and entropy density, respectively, can be written as~\cite{Rischke:1991ke}
\be
n^{EV}(T,\mu)=\sum_{a}\frac{n^{id}_{a}(T,\tilde\mu)}{1+\sum_{a}v_{a}n_{a}^{id}(T,\tilde\mu)},
\ee
\be
\epsilon^{EV}(T,\mu)=\sum_{a}\frac{\epsilon^{id}_{a}(T,\tilde\mu)}{1+\sum_{a}v_{a}n_{a}^{id}(T,\tilde\mu)},
\ee
\be
s^{EV}(T,\mu)=\sum_{a}\frac{s^{id}_{a}(T,\tilde\mu)}{1+\sum_{a}v_{a}n_{a}^{id}(T,\tilde\mu)}.
\ee

Again in the Boltzmann approximation all the thermodynamical quantities are multiplied by the factor $\text{exp}(-vP^{EV}/T)$.
 But unlike pressure there is an additional factor $\frac{1}{1+\sum_{a}v_{a}n_{a}(T,\tilde\mu)}$, which suppresses the thermodynamical quantities at high temperature as compared to their ideal gas counterpart. Once the thermodynamic quantities are estimated, we can calculate the electrical and thermal conductivities using the EHRGM model.

\section{Transport coefficients in relaxation time approximation}
\label{secIII}
\subsection{Electrical conductivity}
The electric conductivity ($ \sigma_{el} $) represents the response of the system to an applied electric field,  
\begin{equation}
\vec{j}=\sigma_{el} \vec{E}.
\end{equation}
We start our calculation from the relativistic Boltzmann transport (RBT) equation. 
In the presence of an external field, the RBT equation can be written as~\cite{Yagi, Cercignani}
\begin{equation} 
\label{Boltzmann_eq}
k^{\mu}\partial_{\mu} f_{a}(x,k) + q_{a} F^{\alpha\beta}k_{\beta} \frac{\partial}{\partial k^{\alpha}} f_{a}(x,k) = {\cal C}_{a}[f_{a}],
\end{equation}
where $F^{\alpha\beta}$ is the electromagnetic field strength 
tensor and  ${\cal C}_{a}[f_{a}]$ is 
the collision integral. Here we have introduced the index $ a $ on the distribution function for the hadronic species. The relaxation time 
approximation (RTA) is the simplest scheme to approximate the collision term ${\cal C}_{a}[f_{a}]$, which is given by 
\begin{equation}
{\cal C}_{a}[f_{a}] \simeq -\frac{k^{\mu}u_{\mu}}{\tau_{a}} \delta f_{a}, 
\label{cf}
\end{equation}
where $ u_{\mu}=(1, {\bf 0}) $ is the fluid four velocity in the local rest frame and $ \tau_{a} $ is the relaxation time, which estimate the timescale for the system to relax towards the equilibrium state.
 $\delta f_{a}= f_{a}-f_{a}^{0}$, where we assume that the  distribution function $ f_{a} $ is very close to the equilibrium distribution $ f_{a}^{0} $ and can be written for deviation in linear order as~\cite{Puglisi:2014sha}
\begin{equation}
f_{a}(x,{\bf k})=f_{a}^{0}(x,{\bf k})(1+\varphi(x,{\bf k}))=f_{a}^{0}+\delta f_{a},
\end{equation}
%
where $ \varphi $ ($ \mid \varphi \mid \ll 1$) is the perturbation. The equilibrium 
particle distribution function is
\begin{equation}
f_{a}^{0}(x,{\bf k})=\frac{1}{e^{(E_{a}\pm \mu_{a})/T}\pm 1}, \, \, \, E_{a}=\sqrt{{\bf k}^{2}+m_{a}^{2}},
\end{equation}
where $ \pm $ corresponds to fermion and boson, respectively. 
For constant electric field $ \bf {E} $,
 Eq. (\ref{Boltzmann_eq}) becomes
\begin{equation}
q_{a}\left( k_0 {\bf E} \cdot \frac{\partial f_{a}^{0}}{\partial {\bf k}} + {\bf E}\cdot {\bf k} \frac{\partial f_{a}^{0}}{\partial k^0} \right) = -\frac{k^0}{\tau_a}\delta f_{a}. 
\label{RBT1}
\end{equation}
After solving one can get $\delta f_{a} $ for the case when  $ \varphi \ll f_{a}^{0} $ as
\begin{equation}
\delta f_{a} = \sum_{a}\frac{q_{a}\tau_{a}}{T}{\bf E} \cdot \frac{ {\bf k}}{k^{0}} f_{a}^{0}(1\pm f_{a}^{0}).
\label{delf}
\end{equation}
The electric four current ($ j^{\mu} $) can be written as
\begin{eqnarray}
j^{\mu}=\sum_{a} q_{a} g_{a}\int \frac{d^{3}k}{(2 \pi)^3E_{a}} k^{\mu}f_{a}(x,k),
\label{current}
\end{eqnarray}
where $ q _{a} $(${\bar q_a} $) and $ f_{a}(x,k)( {\bar f_{a}(x,k)} $) are the charge and distribution 
functions for particles (antiparticles)
$ a $. 
After applying an external disturbance, $ j^{\mu}=j_{0}^{\mu}+ \Delta j^{\mu} $, four current becomes 
\begin{equation}
 \Delta j^{\mu}= q_{a} g_{a} \int \frac{d^{3}k}{(2 \pi)^3E_{a}} k^{\mu} \delta f_{a}.
\label{delj}
\end{equation}
Considering the definition of electrical conductivity and  substituting $ \delta f_{a} $ into that, we get 
\begin{equation}
\sigma_{\rm{el}} = \frac{1}{3T} \sum_{a} g_{a}q_{a}^{2}\int\frac{d^3k}{(2\pi)^3} 
\frac{{k}^2}{E_{a}^2}  \tau_{a} \times f_{a}^{0}(1\pm f_{a}^{0}).
\label{sigiso}
\end{equation}
In the Boltzmann approximation the above equation can be written as
\begin{equation}
\sigma_{\rm{el}} = \frac{1}{3T} \sum_{a} g_{a}q_{a}^{2}\int\frac{d^3k}{(2\pi)^3} 
\frac{{k}^2}{E_{a}^2}  \tau_{a} \times f_{a}^{0}.
\label{sigiso}
\end{equation}
\subsection{Thermal Conductivity}
Thermal conductivity $ \kappa $ is interesting to study, as it describes the heat flow in interacting systems~\cite{Israel:1979wp,Groot}. Recently it has reattained interest in the context of relativistic HICs~\cite{Greif:2013bb,Denicol:2012cn}. We will start our calculations from the RBT equation. In the absence of external field, Eq. (\ref{Boltzmann_eq}) can be written as~\cite{Hosoya:1983xm}
\begin{equation}\label{Boltzmann_eq2}
k^{\mu}\partial_{\mu} f_{a}(x,k) = -\frac{k^{\mu}u_{\mu}}{\tau_a} \delta f_{a}.
\end{equation}
We start our calculation from the energy momentum tensor $ (T^{\mu\nu})$ and four current $ (j^{\mu}) $, which are, respectively, given by~\cite{Hosoya:1983xm,Gavin:1985ph}
\begin{equation}
T^{\mu\nu}=\sum_{a}g_{a}\int \frac{d^3k}{(2\pi)^3E_{a}}k^{\mu}f_{a}(x,k)
\end{equation}
and 
\begin{equation}
j^{\mu}=\sum_{a}g_{a}\int \frac{d^3k}{(2\pi)^3E_{a}}k^{\mu}k^{\nu}f_{a}(x,k),
\end{equation}
where, as before, $ t_{a} $ and $ g_{a} $ are , respectively, the charge and the degeneracy of hadronic
species a.
In the presence of a small disturbance from the equilibrium distribution function, 
the change in energy momentum tensor $ \Delta T^{\mu\nu} $ can be written as
\begin{equation}
\Delta T^{\mu\nu}=\sum_{a}g_{a}\int \frac{d^3k}{(2\pi)^3E_{a}}k^{\mu}k^{\nu}\delta f_{a}(x,k).
\label{deltaT}
\end{equation}
%
Using the RTA, $ \Delta T^{\mu\nu} $ becomes~\cite{Hosoya:1983xm}
\begin{eqnarray}
\Delta T^{\mu\nu}=-\sum_{a}g_{a}\int \frac{d^3k}{(2\pi)^3E_{a}}\frac{k^{\mu}k^{\nu}}{k.u} \tau_{a}k^{\alpha}\partial_{\alpha}f_{a}(x,k)
\end{eqnarray}
and the change in four current $ \Delta j^{\mu} $ becomes
\begin{equation}
\Delta j^{\mu}= \sum_{a} g_{a} \int \frac{d^{3}k}{(2 \pi)^3E_{a}}\frac {k^{\mu}}{k.u} \tau_{a}k^{\alpha}\partial_{\alpha}f_{a}(x,k),
\label{delj1}
\end{equation}
where $ \partial_{\mu}=u_{\mu}D+\nabla_{\mu} $, and the convective derivatives $(DT, D\mu, Du^{\mu})$ can be eliminated by using the relation
\begin{eqnarray}\label{1a}
(\varepsilon+P)Du^{\mu}-\nabla^{\mu}P=0, \\
Dn+n\nabla_{\mu}u^{\mu}=0.
\end{eqnarray}
After using the above relations, one can obtain~\cite{Hosoya:1983xm}      
\begin{eqnarray}
\Delta T^{\mu\nu}&=&\sum_{a}g_{a}\int \frac{d^3k}{(2\pi)^3E_{a}}\frac{k^{\mu}k^{\nu}}{k.u}\frac{1}{T}\Bigg[ \tau_{a}f_{a}^{0}(1-f_{a}^{0}) 
\bigg\{k.u{\left(\frac{\partial k}{\partial \varepsilon}\right)}_{n}\nabla_{\alpha}u^{\alpha}
+ k^{\alpha}X_{\alpha}+\frac{k^{\alpha}k^{\beta}}{k.u} \nabla_{\alpha}u_{\beta}\nonumber\\
&+&\left(\frac{\partial k}{\partial n}\right)_{\varepsilon}\nabla_{\alpha}u^{\alpha}-\frac{\varepsilon+P}{n}\frac{k^{\alpha}}{k.u}X_{\alpha}\bigg\}\Bigg],
\end{eqnarray}
and
\begin{eqnarray}
\Delta j^{\mu}&=&\sum_{a}g_{a}\int \frac{d^3k}{(2\pi)^3E_{a}}\frac{k^{\mu}}{k.u}\frac{1}{T}\Bigg[ \tau_{a}f_{a}^{0}(1-f_{a}^{0}) 
\bigg\{k.u{\left(\frac{\partial k}{\partial \varepsilon}\right)}_{n}\nabla_{\alpha}u^{\alpha}
+ k^{\alpha}X_{\alpha}+\frac{k^{\alpha}k^{\beta}}{k.u} \nabla_{\alpha}u_{\beta}\nonumber\\
&+&\left(\frac{\partial k}{\partial n}\right)_{\varepsilon}\nabla_{\alpha}u^{\alpha}-\frac{\varepsilon+P}{n}\frac{k^{\alpha}}{k.u}X_{\alpha}\bigg\}\Bigg],
\end{eqnarray}
where
\begin{equation}
X_{\alpha}=\frac{\nabla_{\alpha}P}{\varepsilon+P}-\frac{\nabla_{\alpha}T}{T},
\end{equation}
and $ u_{\mu}=(1,\bfm{0})$. $ \varepsilon $ and $ n $ are the energy density and number density. 
The momentum conservation shows that $ \bfm{\nabla P}=0$ [where $\bfm{\nabla P}=(\varepsilon+P)\partial u/\partial t]$ in the steady state.
Thermal conduction, which involves the relative flow of energy, which arises when energy flows relative to the baryonic enthalpy. The $ T^{0i} $ component is the energy flux and with the Eckart condition, $ T^{0i}=\Delta T^{0i}-\frac{(\varepsilon+P)}{n}\Delta j^{i}\equiv I^{i} $,
where $ I^{i} $ is the heat current with
\begin{eqnarray}\label{T0i}
\Delta T^{0i}&=&\sum_{a}g_{a}\int \frac{d^3k}{(2\pi)^3}\frac{ \bfm{k^2}}{3T}\tau_{a}f_{a}^{0}(1-f_{a}^{0})\bigg\{1-\frac{\varepsilon+P}{nE_a}\bigg\}X_{i}.
\end{eqnarray}
and 
\begin{eqnarray}\label{ji}
\Delta j^{i}&=&\sum_{a}g_{a}\int \frac{d^3k}{(2\pi)^3E_{a}}\frac{ \bfm{k^2}}{3T}\tau_{a}f_{a}^{0}(1-f_{a}^{0})\bigg\{1-\frac{\varepsilon+P}{nE_a}\bigg\}X_{i}.
\end{eqnarray}
Using either the Eckart or Landau-Lifshitz condition, one can define the heat conductivity as~\cite{Hosoya:1983xm}
\begin{equation}\label{key1}
I^{i}=-\kappa\left[\partial_{i}T-T\partial_{i}P/(\varepsilon+P)\right]=\kappa T X_{i}.
\end{equation}
Using Eqs. (\ref{T0i}) and (\ref{ji}), one can obtain the expression for thermal conductivity as
\begin{eqnarray}\label{thermalcond}
\kappa=\frac{1}{3T^{2}}\sum_{a}g_{a}\tau_{a}\int \frac{d^3k}{(2\pi)^3}\frac{\bf k^2}{E_{a}^2}f_{a}^{0}(1-f_{a}^{0})\left(E_{a}-\frac{t_a\omega}{n}\right)^{2},
\end{eqnarray}
where $ \omega=\varepsilon+P $ is the enthalpy and $ t_a=+1(-1) $ for particles (anti-particles). For the baryonic matter and low temperature, the antiparticle contribution can be neglected, as the temperature are much smaller than the masses of the baryon. Since we will work in the Boltzmann approximation, the expression for thermal conductivity can be written as
\begin{eqnarray}\label{thermalcond}
\kappa=\frac{1}{3T^{2}}\sum_{a}g_{a}\tau_{a}\int \frac{d^3k}{(2\pi)^3}\frac{\bf k^2}{E_{a}^2}f_{a}^{0}\left(E_{a}-\frac{t_a\omega}{n}\right)^{2}.
\end{eqnarray}
%
\section{Relaxation Time}
\label{secIV}
The relaxation time $ \tau_{a} $ is defined by the expression~\cite{Kadam:2015xsa}
\begin{equation}
\tau^{-1}(E_{a})=\sum_{bcd}\int\frac{d^{3}p_{b}}{(2\pi)^{3}}\frac{d^{3}p_{c}}{(2\pi)^{3}}\frac{d^{3}p_{d}}{(2\pi)^{3}}W(a,b\rightarrow c,d)f_{b}^{0},
\label{tau}
\end{equation}
where $W(a,b\rightarrow c,d)$ is the transition rate,
\begin{equation}
W(a,b\rightarrow c,d)=\frac{(2\pi)^{4}\delta(p_{a}+p_{b}-p_{c}-p_{d})}{2E_{a}2E_{b}2E_{c}2E_{d}}\mid \mathcal{M}\mid^{2},
\end{equation}
and $\mid \mathcal{M}\mid$ is the transition amplitude.
Equation (\ref{tau}) can be simplified in the center of mass frame as
\be
\tau^{-1}(E_{a})=\sum_{b}\int\frac{d^{3}p_{b}}{(2\pi)^{3}}\sigma_{ab}v_{ab}f_{b}^{0},
\label{relx}
\ee
where $\sigma_{ab}$ is the total scattering cross section for the process, $a(p_{a})+b(p_{b})\rightarrow a(p_{c})+b(p_{d})$, and $v_{ab}$ is relativistic relative velocity. One can obtain the averaged partial relaxation time by averaging the relaxation time over $f_{a}^{0}$, which is rather a good approximation~\cite{Moroz:2013haa}. Thus, the averaged relaxation time can be written as
\be
{\tilde\tau}_{a}^{-1}=\sum_{b}n_{b}\langle\sigma_{ab}v_{ab}\rangle.
\label{relxaverage}
\ee
 In the above, $v_{ab}$ is the ``relative velocity" defined by
$$
v_{ab}=\frac{\sqrt{(p_a\cdot p_b)^2-m_a^2m_b^2}}{E_aE_b}
$$
with $p_a$, $E_a$ being the four momentum and energy of particle ``a" and, with $g_b$ being the degeneracy of species ``b",
$$n_{b} =\frac{g_b}{(2\pi)^3}\int\frac{d^{3}p_{b}}{(2\pi)^{3}}f_{b}^{0}$$ 
is the  equllibrium number density of  $b$th hadronic species. Here we use the equilibrium Maxwell-Boltzmann distribution 
\be
f_{a}^{0}=exp\bigg(-\frac{E_{a}-\mu_a}{T}\bigg).
\label{MB}
\ee
The thermal average of total cross section times relative velocity, i.e., $\langle\sigma v\rangle$ for the scattering of hard 
sphere particles of the same species at  a given $T$ and $\mu$ (having constant cross section, $\sigma$), can be calculated 
as outlined in Refs.~\cite{Cannoni:2013bza,Gondolo:1990dk}.
The thermal average $\langle\sigma v\rangle$ for the process $a(p_{a})+a(p_{b})\rightarrow a(p_{c})+a(p_{d})$ can be written as
\be
\langle\sigma_{ab} v_{ab}\rangle=\frac{\sigma \int d^{3}p_{a}d^{3}p_{b} v_{ab}e^{-E_{a}/T}e^{-E_{b}/T}}{\int d^{3}p_{a}d^{3}p_{b}e^{-E_{a}/T}e^{-E_{b}/T}}.
\label{thermalave}
\ee
Note that, in the above, the chemical potential dependences gets canceled from numerator and the denominator, which is a
consequence of Boltzmann approximtions for the equilibrium thermal distribution function.
After changing the integration variable as discussed in detail in Ref.~\cite{Kadam:2015xsa}. The numerator and denominator in Eq. (\ref{thermalave}) becomes
\be
\int d^{3}p_{a}d^{3}p_{b} v_{ab}e^{-E_{a}/T}e^{-E_{b}/T}=2\pi^{2}T\int ds \sqrt{s} (s-4m^{2})K_{1}(\sqrt{s}/T),
\ee
and
\be
\int d^{3}p_{a}d^{3}p_{b}e^{-E_{a}/T}e^{-E_{b}/T}=[4\pi m^{2}T K_{2}(m/T)]^{2}.
\ee
Therefore, the thermal average $ \langle\sigma_{ab}v_{ab}\rangle $ can be written as~\cite{Kadam:2015xsa}
\be
\langle\sigma_{ab} v_{ab}\rangle=\frac{\sigma}{8m^{4}TK_{2}^{2}(m/T)}\int_{4m^{2}}^{\infty}ds\sqrt{s}(s-4m^{2})K_{1}(\sqrt{s}/T),
\label{thermal1}
\ee
where $\sqrt{s}$ is the center of mass energy and  $K_{1}$ ($K_{2}$) is the modified Bessel function of order $ 1 (2) $. For the case of scattering between different species of the particles, 
Eq. (\ref{thermal1}) becomes
\be
\langle\sigma_{ab}v_{ab}\rangle=\frac{\sigma}{8Tm_{a}^{2}m_{b}^{2}K_{2}(\frac{m_{a}}{T})K_{2}(\frac{m_{b}}{T})}\int_{m_{a}+m_{b}}^{\infty}ds\frac{[s-(m_{a}-m_{b})^{2}]}{\surd s}[s-(m_{a}+m_{b})^{2}]K_{1}(\surd s/T).
\label{thermalave2}
\ee
After evaluating the thermal averaged cross section, we can relate it to the relaxation time in Eq. (\ref{relxaverage}). 

\section{Results and discussion}
\label{secV}
In the hadron resonance gas model, it is customary to include all the hadrons and resonances up to certain cutoff $\Lambda$. We choose cutoff $\Lambda=2.25$ GeV and include all the mesons and baryons listed in Ref.~\cite{Amsler:2008zzb}. The only parameter that remains in our model is hard-core radius $r_{h}$ or the excluded volume parameter $v$. We choose two different parametrization schemes, viz., uniform excluded volume parameter $(v=\frac{16}{3}\pi r_{h}^{3})$~\cite{Gorenstein:2007mw} and mass dependent excluded volume parameter ($v=\frac{M_{h}}{\epsilon_{0}}$)~\cite{Kapusta:1982qd}. Here $\epsilon_0$ is the parameter that we fix to 2 GeV fm$^{-3}$. Based on the the nucleon-nucleon scattering analysis\cite{bohr} we choose uniform hard-core radius $r_h=0.3$fm.\\

\begin{figure}[t]
	\vspace{-0.4cm}
	\begin{center}
		\begin{tabular}{c c}
			\includegraphics[width=9cm,height=9cm]{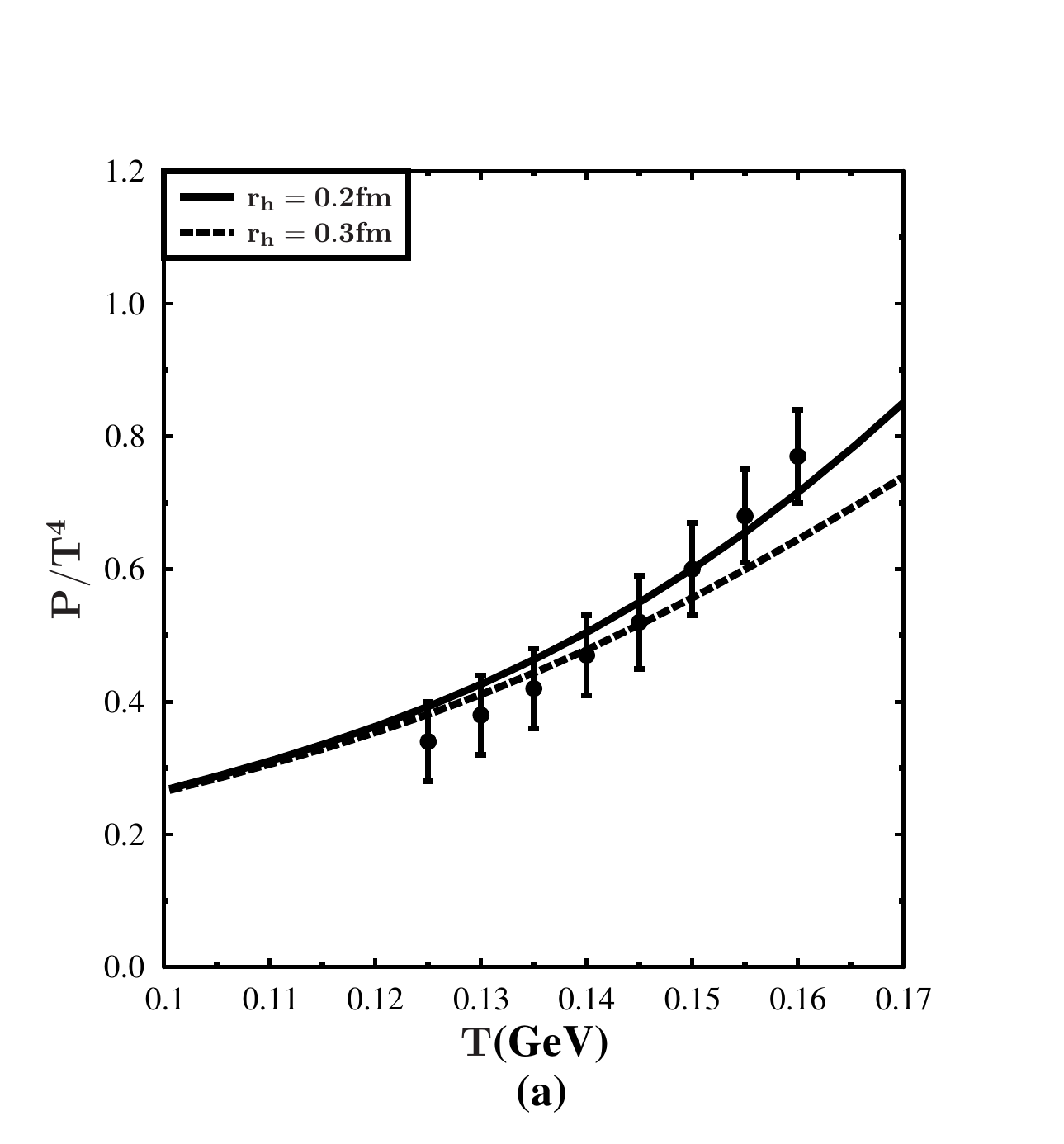}&
			\includegraphics[width=9cm,height=9cm]{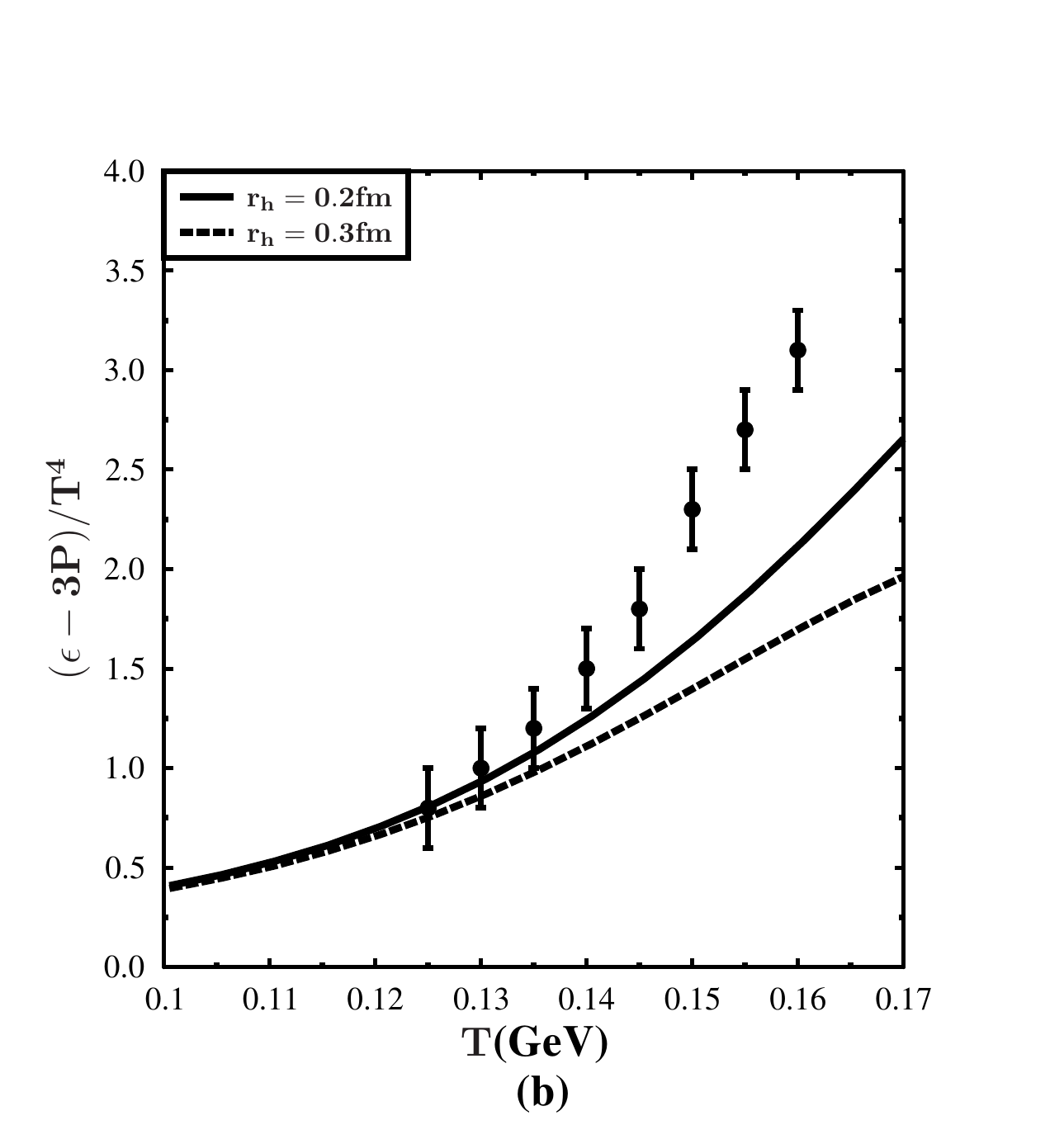}
		\end{tabular}
		\caption{ Thermodynamical functions, pressure (a) and trace anomaly (b), at zero chemical potential.} 
		\label{thermu0}
	\end{center}
\end{figure}

Figure~\ref{thermu0} shows the thermodynamical properties of hadron gas estimated within the ambit of the EHRGM. Figure~\ref{thermu0}(a) shows scaled pressure $P/T^4$ at zero baryon chemical potential for two different choices of hadron hard-core radius $r_{h}=0.2$ and $r_{h}=0.3$fm. We note that EHRGM estimates deviate from the lattice data at higher temperature. The deviation is large for larger hard-core radius. This is essentially due to the suppression factor $(1+vn^{EV})^{-1}$ which is large for higher $r_{h}$. Fig.\ref{thermu0}(b) shows the scaled interaction measure $(\epsilon-3p)/T^4$. Again the EHRGM estimates strongly deviate from the lattice data at higher temperature. The rapid rise in the trace anomaly cannot be explained within EHRGM model alone. But it has been shown in Ref.\cite{Vovchenko:2014pka} that by including the Hagedorn mass spectrum along with the discrete hadron spectrum in the HRG model the resulting excluded volume model reproduces the lattice data up to $160$MeV at $\mu=0$. Similar studies extended to include finite baryon chemical potential confirm this result\cite{Kadam:2018hdo}.  Note that we will not include the Hagedorn states in our calculations since their quantum numbers, especially the electric charges are not known experimentally.

\begin{figure}[h]
	\vspace{-0.4cm}
	\begin{center}
		\includegraphics[width=10cm,height=10cm]{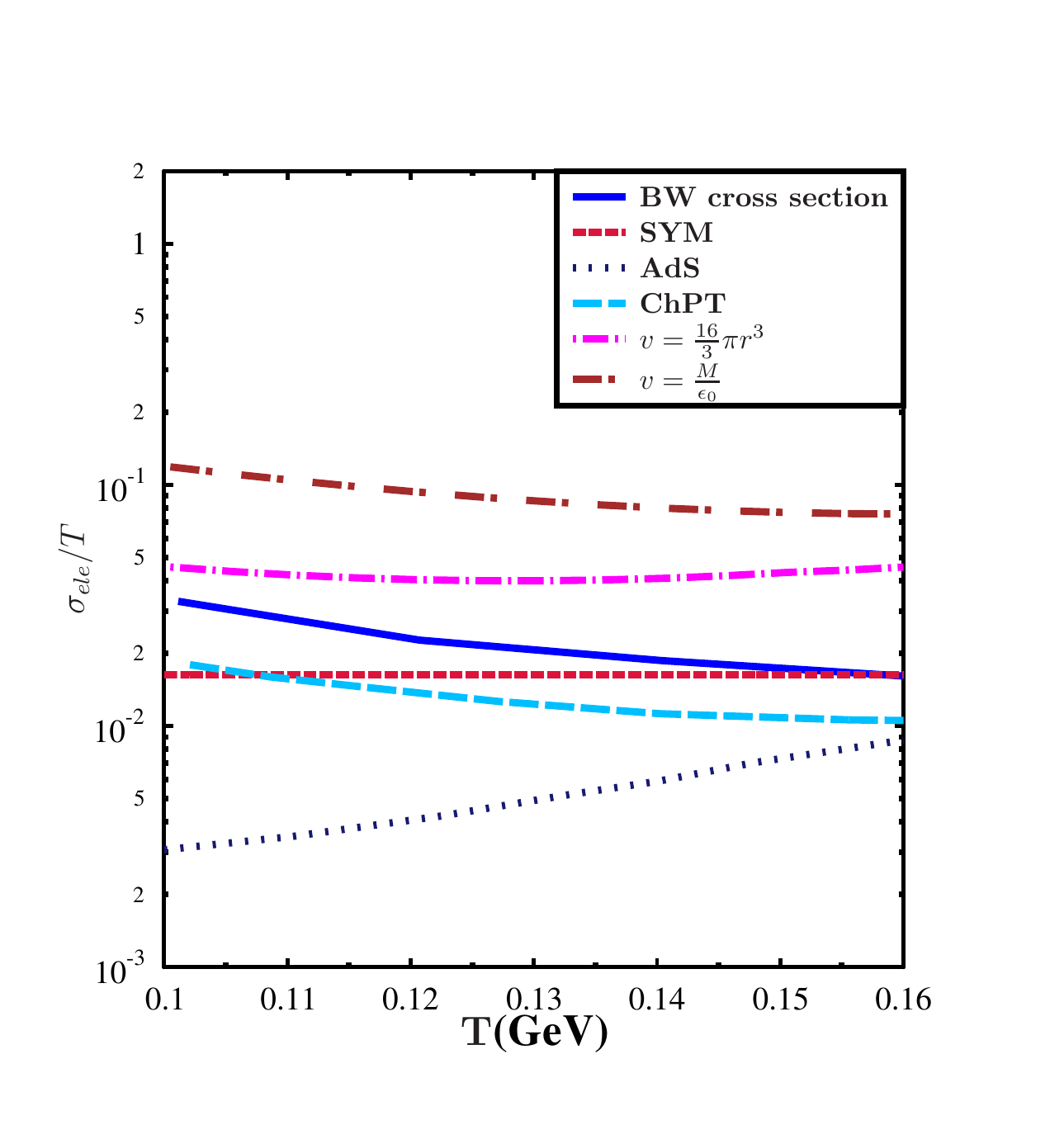}
		\caption{Normalized conductivity in the present the EHRGM with two different
			parametrizations for the excluded volume compared with other model estimations. The magenta  dot-dashed curve
			refers to EHRGM with a uniform  excluded volume parameter while the dashed double-dot curve corresponds to the EHRGM with a mass
			dependent excluded volume calculation.} 
		\label{econd_comp}
	\end{center}
\end{figure}
In Fig.~\ref{econd_comp}, we show the dimensionless electrical conductivity  ($ \sigma_{ele}/T $) as a function of temperature at zero chemical potential. We have compared our results with the various results that exist in the literature. The red dashed line shows the results of the conformal Super-Yang-Mills (SYM) plasma~\cite{Starinets2006}.
The 
red open circles represent the data
from lattice QCD calculation~\cite{Aarts:2014JHEP}. However, the hadronic interactions are missing in the lattice calculation. The violet dotted line represents the  nonconformal holographic model~\cite{Rougemont:2015ona, Finazzo:2013efa}.
The cyan dashed line represents  chiral perturbation theory (CPT) results~\cite{Fernandez-Fraile2006}. The blue solid line shows the kinetic theory results~\cite{Greif:2016skc}. The magenta curve shows our results for the uniform excluded volume parameter $(v=\frac{16}{3}\pi r_{h}^{3})$,  while the maroon curve corresponds to the mass dependent excluded volume parameter($v=\frac{M_{h}}{\epsilon_{0}}$).  The behavior of $ \sigma_{ele}/T $ with temperature from the CPT and kinetic theory results are similar to our results, although there is a difference in magnitude of electrical conductivity.  The magnitude of electrical conductivity is higher in the model as compared to other results, especially the kinetic theory estimations of Ref.~\cite{Greif:2016skc}. However, this is not so surprising. The basic reason behind higher conductivity in our model is the smaller cross section. In the case of uniform excluded volume parameter the cross sections $\sim 10$mb for all the hadronic species, while in Ref. ~\cite{Greif:2016skc} different cross sections are assumed for different species and the values of the cross section are relatively large. Since the conductivity is inversely proportional to the cross section (through relaxation time $\tau$), its estimation turns out to be large in our model. However, it may be noted that assigning hard-core radius to all the hadrons may not be the correct way to  account for the repulsive interactions within the noninteracting HRG model. One possible improvement one can do to this model is to assign repulsive interactions only between baryons and anti-baryons while mesons are kept non-interacting~\cite{Vovchenko:2016rkn}.  Estimating the transport coefficients within this model is under progress and will appear elsewhere.  

\begin{figure}[t]
	\vspace{-0.4cm}
	\begin{center}
		\begin{tabular}{c c}
			\includegraphics[width=9cm,height=9cm]{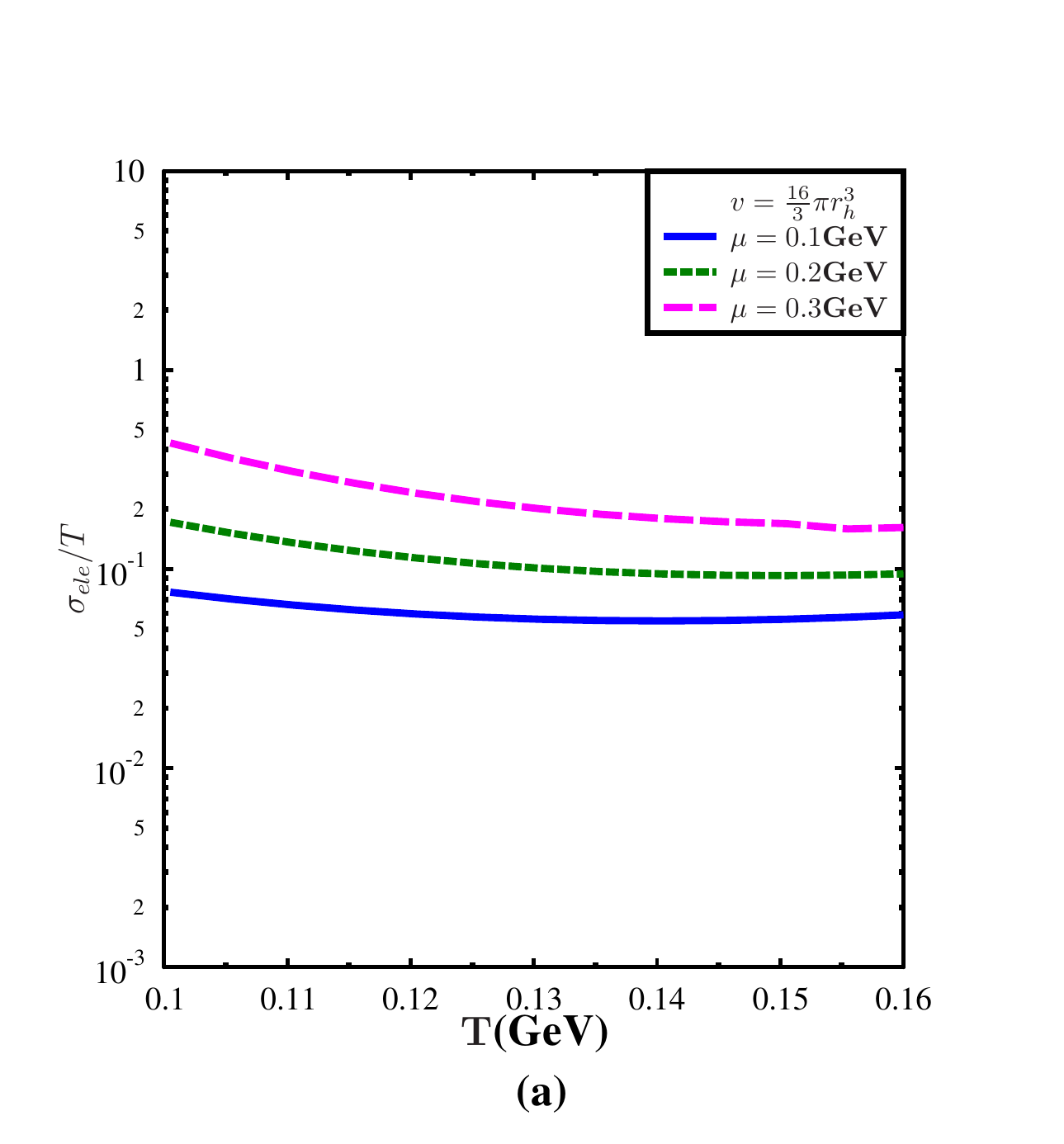}\label{ele_condaa}&
			\includegraphics[width=9cm,height=9cm]{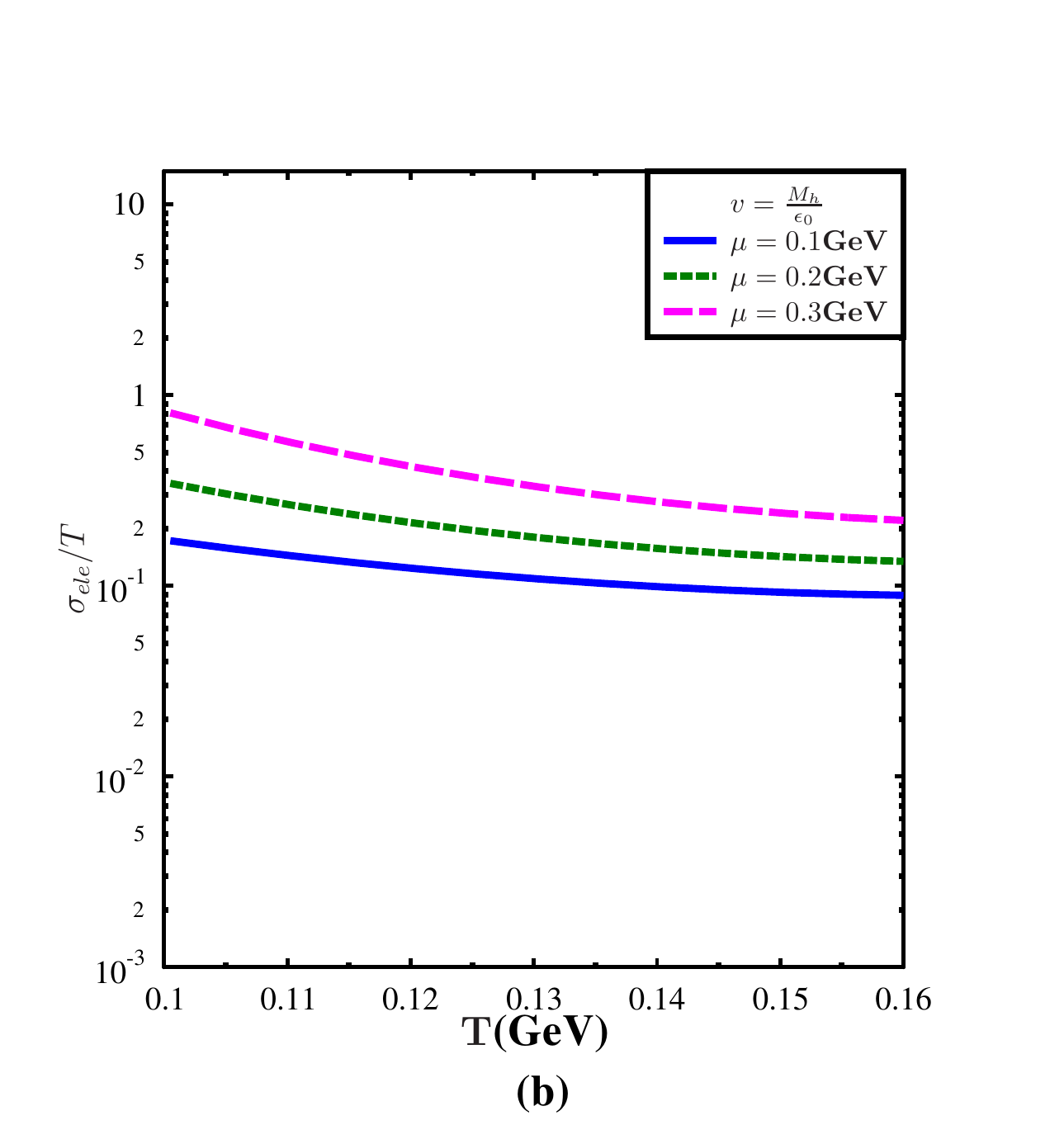}\label{ele_condbb}
		\end{tabular}
		\caption{ The scaled electrical conductivity as a function of temperature for different values for the baryon chemical potential.
			(a) shows the variation in the EHRGM with a uniform hard-core excluded volume. (b) Corresponds to EHRGM model with a mass dependent excluded volume.}
		\label{ele_cond}
	\end{center}
\end{figure} 

In Fig.~\ref{ele_cond} we show the variation of electrical conductivity with temperature for different chemical potentials, 
$ \mu= 0.1 $, $ 0.2 $ and $ 0.3 $ GeV. Figure ~\ref{ele_cond}a corresponds to uniform excluded volume parameter, while Fig. ~\ref{ele_cond}b 
corresponds to mass dependent excluded volume parameter. We note that the electrical conductivity increases with the increase
in chemical potential although the general behavior as a function of temperature does not change. This behavior is not hard to 
understand. In Eq. (40), while the cross section is independent of  both $\mu$ and T, the thermally averaged cross section
times the relative velocity $\langle\sigma v\rangle$
is, in general, dependent on both T and $\mu$ arising from the distribution functions.  However, in the Boltzmann approximation,
the $\mu$ dependence gets canceled from the numerator and the denominator.
On the other hand,  for the thermal averaged  cross section times the relative velocity 
or the inverse of scattering length, given by $\langle \sigma v\rangle _{ab}n_b$  (Eq. 38), will be an increasing function 
of $\mu$ if species 'b' is a baryon.
So this will lead to the relaxation time being
a decreasing function of $\mu$. In the expression for the $\sigma_{ele}$, $\tau_a$ is multiplied by a distribution function which again is an increasing function of $\mu$, when species 'a' is a baryon. Thus, the  contribution to $\sigma_{ele}$ from,  say, a baryon will depend upon which of the two parts $\tau_a$ (a decreasing function of $\mu$) and the distribution function $f_{0}$  ( which is an increasing function of $\mu$)
dominate the variation with $\mu$. It turns out that, for baryons, the dominant contribution to the relaxation time
arises from the baryon scattering with mesons and in that case the corresponding avg cross section or, equivalently, the
relaxation time is independent of $\mu$. Therefore, the $\mu$ variation of the corresponding contribution
of the baryon to $\sigma_{ele}$  is an increasing function of $\mu$.

From the figure it is clear that the $\sigma_{ele}/T$ ratio is higher for the mass dependent excluded volume parameter case as compared to the uniform excluded volume parameter case for the different values of chemical potential  although the general behavior of the ratio  is similar as a function of temperature. This behavior may be the reflection of the fact that while the cross section in case of uniform hard core excluded volume parameter is $\sim 10$mb, the same in mass dependent parametrization varies from $3$mb for pions to $10$mb for protons. Thus, the smaller cross section range leads to larger relaxation time with larger electrical conductivity.
\begin{figure}[t]
	\vspace{-0.4cm}
	\begin{center}
		\begin{tabular}{c c}
			\includegraphics[width=9cm,height=9cm]{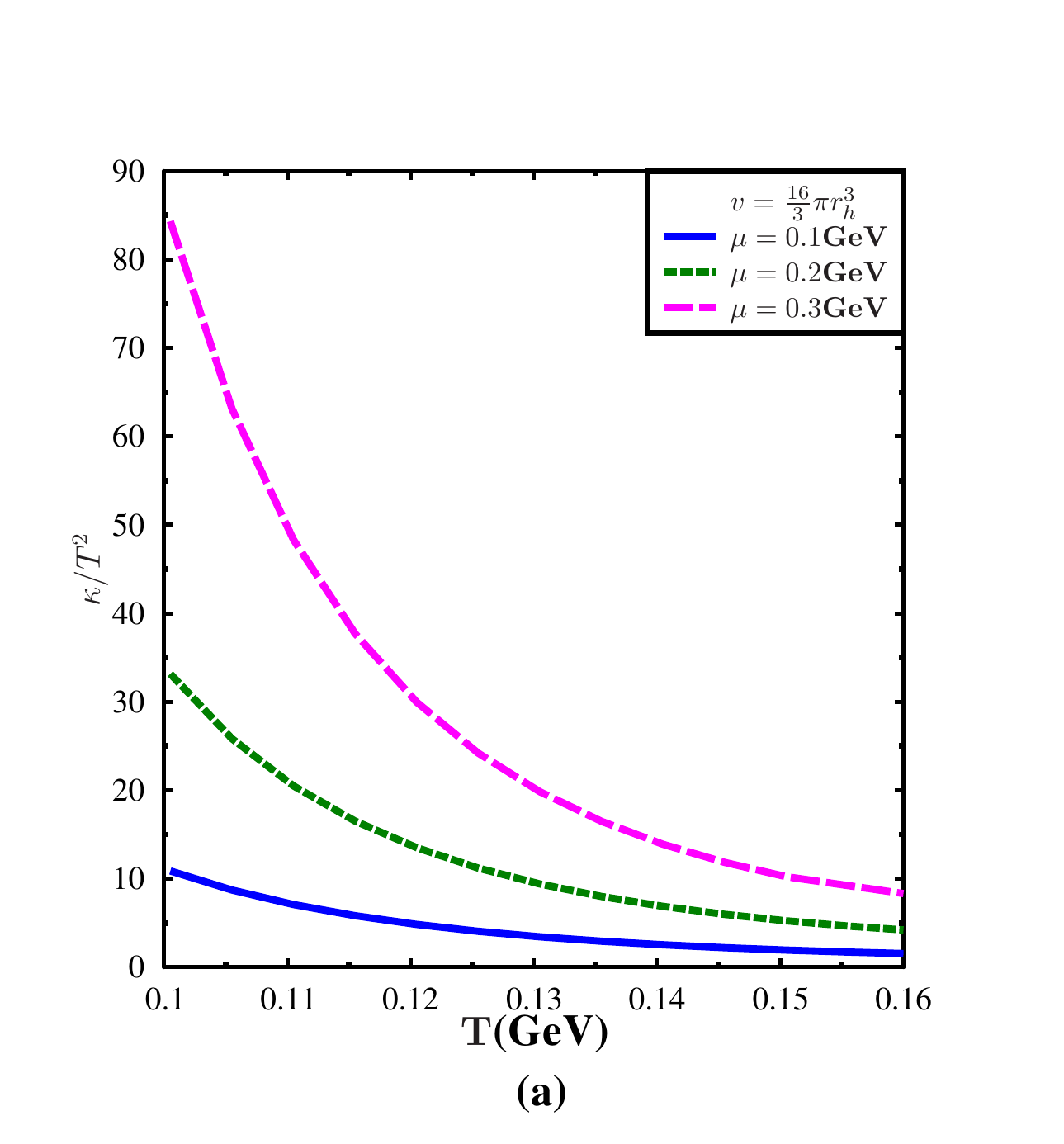}&
			\includegraphics[width=9cm,height=9cm]{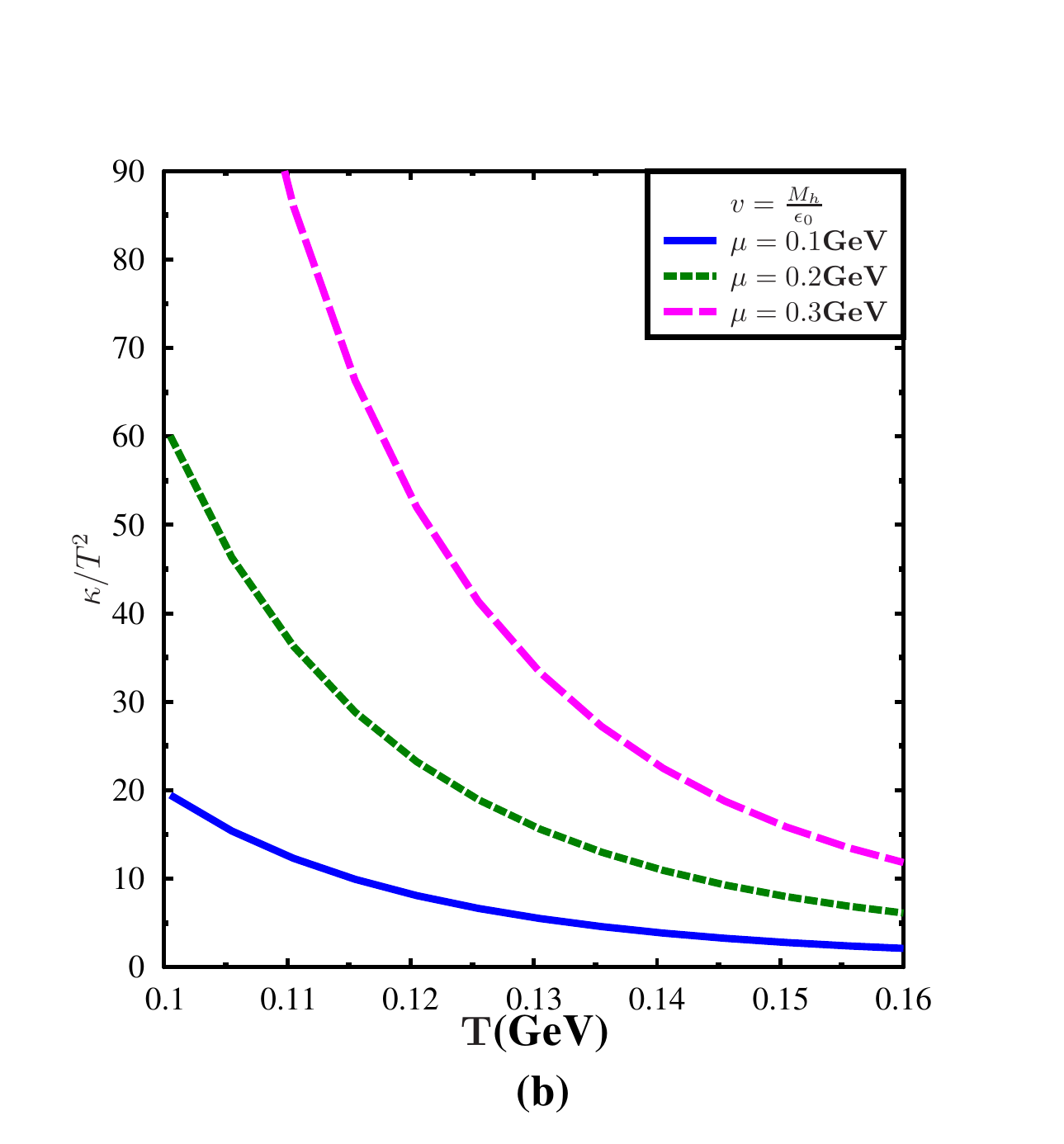}
		\end{tabular}
		\caption{ Scaled thermal conductivity  as a function of temperature in EHRGM with uniform excluded volume for all hadrons (a)
			and a mass dependent excluded volume (b).}
		\label{tcond}
	\end{center}
\end{figure} 

In Fig.~\ref{tcond} we show the variation of thermal conductivity with temperature for $ \mu= 0.1 $, $ 0.2 $, and $ 0.3 $ GeV. We note that the thermal conductivity  decreases with increase in temperature. Further, at a given temperature $\kappa/T^2$ is always larger for higher chemical potential. The coefficient of thermal conductivity depends on three factors, $viz.$ the relaxation time $\tilde{\tau}$, the distribution function $f_0$, and the quantity $w/n$ [see Eq. 34]. Although the relaxation time decreases with $\mu$, $f_0$ and $w/n$ increases with an increase in chemical potential. It turns out that the latter wins over the former and the overall effect is to increase $\kappa$ with $\mu$. We further note that the magnitude of $\kappa/T^2$ in the uniform excluded volume scheme is smaller than that of the mass dependent one. This observation can again be attributed to the fact that the cross section in the former parametrization is relatively larger than that of the latter.

\begin{figure}[t]
	\vspace{-0.4cm}
	\begin{center}
		\begin{tabular}{c c}
			\includegraphics[width=9cm,height=9cm]{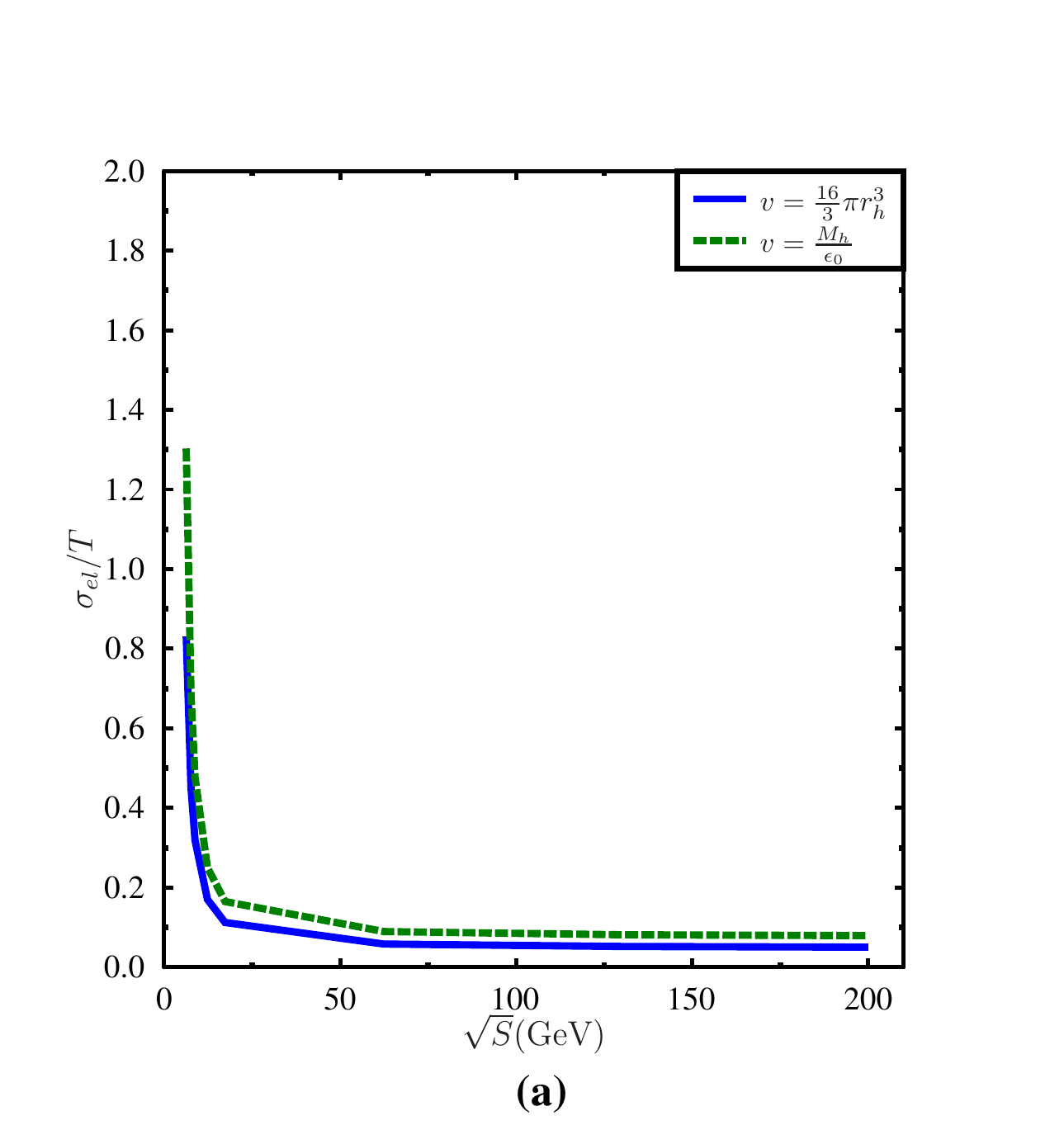}&
			\includegraphics[width=9cm,height=9cm]{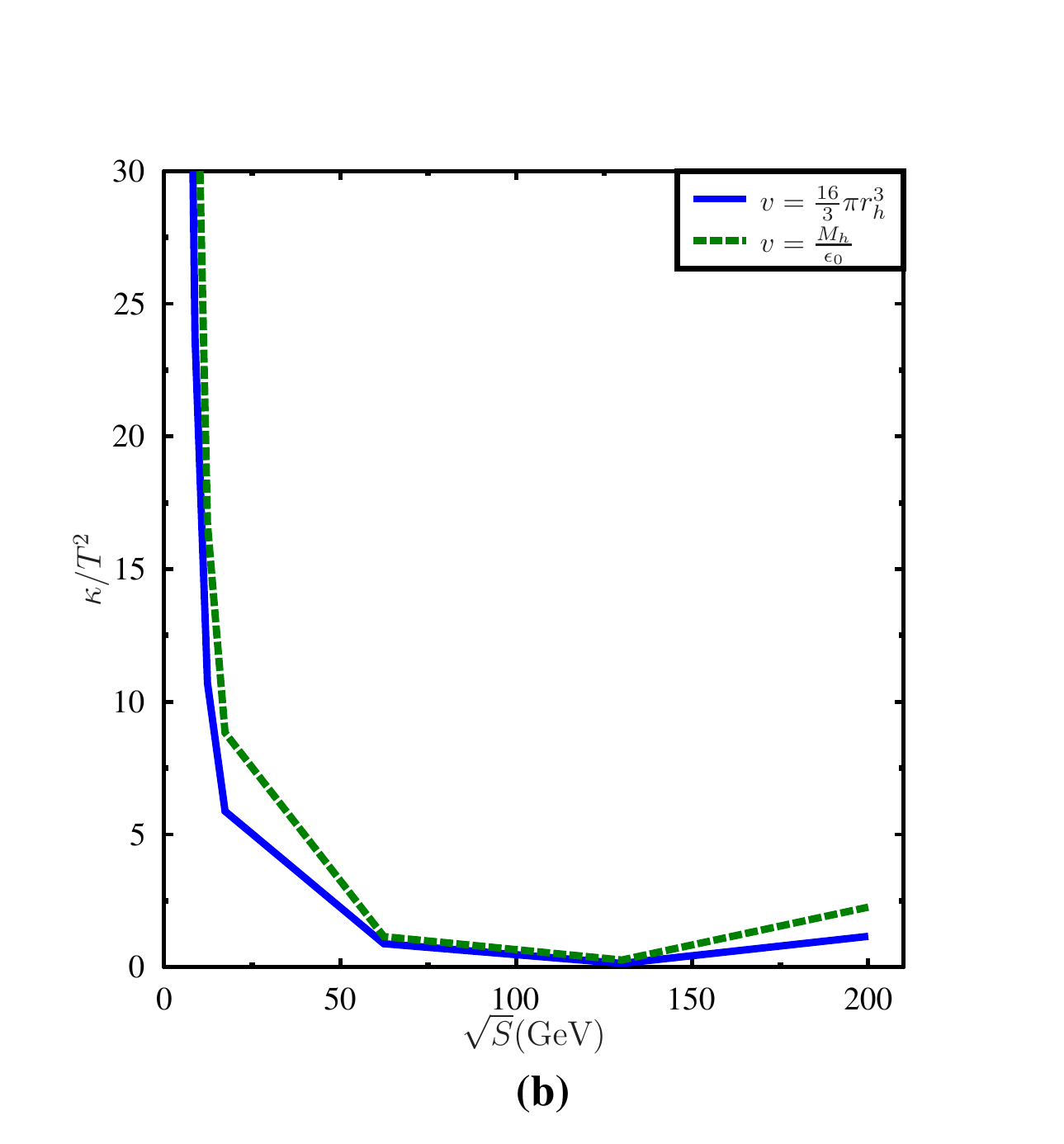}
		\end{tabular}
		\caption{ Variation of scaled electrical conductivity ($\frac{\sigma_{ele}}{T}$) (a) and scaled thermal conductivity ($\frac{\kappa}{T^2}$)
			(b) with the center of mass energy for both uniform excluded volume parameter (blue solid line) and mass dependent volume
			parameter (green dashed line).}
		\label{cme}
	\end{center}
\end{figure}

In order to make the connection with the heavy ion collision experiments we need the beam energy dependence ($\sqrt{s}$) of the electrical and thermal conductivities.  This is extracted from a statistical thermal model description of the particle yield at various $\sqrt{s}$~\cite{Cleymans:2005xv}.  $T(\mu)$ is parametrized by $T(\mu)=a-b\mu^2-c\mu^4$, with $a=0.166\pm0.002$ GeV, $b=0.139\pm0.016$ GeV$^{-1}$ and $c=0.053\pm0.021$ GeV$^{-3}$. The energy dependence of the baryon chemical potential is parametrized as $\mu=d/(1+e\sqrt{s})$, where, $d=1.308\pm0.028$ GeV, and $e=0.273\pm0.008$ GeV$^{-1}$~\cite{Cleymans:2005xv}.
In the Fig.~\ref{cme}a, we have shown the variation of electrical ($ \sigma_{ele}/T $) conductivity with the center of mass energy ($\sqrt{s}$). We note that the electrical conductivity first decreases  along the freeze-out line with increasing collision energy and then attains almost constant value at large $\sqrt{s}$ for both the uniform excluded volume parameter ($v=\frac{16}{3}\pi r_{h}^{3}$) and mass dependent excluded volume parameter case ($v=\frac{M_{h}}{\epsilon_{0}}$). 
This is reasonable because low $\sqrt{s}$ corresponds to low temperature and high chemical potential along the freeze out curve at which electrical conductivity is larger. We can conclude that along the freeze-out line electrical conductivity  of the hadron gas does not change. 

In Fig.~\ref{cme}b, we have shown the variation of thermal conductivity ($ \kappa/T^2 $) with $\sqrt{s}$. We observe that thermal conductivity first decreases  at small values of $\sqrt{s}$ then decreases slowly, becomes minimum and finally increases at its larger values for both the uniform excluded volume parameter and mass dependent excluded volume parameter case.
This can be understood from the expression for thermal conductivity as given in Eq.~(\ref{thermalcond}). In order to discuss the Fig.~\ref{cme}b, we can approximate $ \mu $, $ T $ to be much smaller than the masses of the baryons. In that case we have $  \omega=\epsilon+p\simeq n(m+T) $,
where the baryon number density is given by
\begin{equation}\label{key}
n=2g\left(\frac{mT}{2\pi}\right)^{3/2}e^{-\beta m}\sinh(\beta\mu)
\end{equation}
so that the factor $ (E_a-\frac{\omega}{n})^2$ in Eq.~(\ref{thermalcond}) becomes $ \simeq (E_a-\frac{(m+T)T}{2\mu})^2$.
Therefore as $ \mu $ increases the second term in the parentheses decreases, leading to an increase of the thermal conductivity as seen in Figs.~(\ref{tcond}) and (\ref{cme}). However, as $ \mu $ becomes vanishingly small, the second term dominates over the first term and diverges for $ \mu=0 $. Therefore, $ \kappa/T^2 $ will show a minimum as a function of $\sqrt{s}$ as seen in Fig.~(\ref{cme}b). 
Similar to the electrical conductivity, the value of $ \kappa/T^2 $ is also more with the mass dependent excluded volume parameter case as compared to the case of the uniform excluded volume parameter for all the values of $\sqrt{s}$ apart for its small values.
\section{summary}
\label{secVI}
We have studied the electrical and thermal conductivity of hot and dense hadron gas by using the Boltzmann equation in the relaxation time approximation. First we have estimated the relaxation time for all the hadrons by assuming the constant cross section. 
Here we have used the hadron resonance gas model where the repulsive interactions are parametrized through excluded volume corrections in the ideal hadron resonance gas. We choose the uniform excluded volume and mass dependent excluded volume parametrization scheme. We have included all the hadrons and their resonances with mass cutoff 2.25 GeV. 
Here we take $ r_h=0.3 $ fm for hadrons.
We have compared our results for both the mass dependent excluded volume parameter and the uniform excluded volume parameter case with various existing results. We found that the magnitude of electrical conductivity is higher in our case as compare to these existing results and is more for 
mass dependent excluded volume parameter case as compared to the case of uniform excluded volume parameter.  
We have shown the behavior of electrical and thermal conductivity with temperature for different values of the chemical potential. We found that the electrical and thermal conductivity increases with increase in the chemical potential. The increase in electrical and thermal conductivity is more for the mass dependent excluded volume parameter case as compared to the case of the uniform excluded volume parameter.  %

Further, 
we have shown the variation of electrical ( $ \sigma_{ele}/T $)  and thermal ($ \kappa/T^2 $) conductivity with the collision energy ($\sqrt{s}$). We found that electrical conductivity first decreases  at small values of $\sqrt{s}$ and then remains almost constant at its larger value for both the case of the uniform excluded volume parameter and the mass dependent excluded volume parameter. Thus, we can conclude that electric conductive behavior of hadrons remains same along the freeze-out line. On the other hand, thermal conductivity first decreases  with $\sqrt{s}$, attains minimum, and then increases very slowly.
\section{Acknowledgement}
G.K. is financially supported by the DST-INSPIRE faculty award under Grant No.DST/INSPIRE/04/2017/002293.

\end{document}